\documentstyle[12pt]{article}
\catcode`\@=11

\def\@normalsize{\@setsize\normalsize{15pt}\xiipt\@xiipt
\abovedisplayskip 14pt plus3pt minus3pt%
\belowdisplayskip \abovedisplayskip
\abovedisplayshortskip  \z@ plus3pt%
\belowdisplayshortskip  7pt plus3.5pt minus0pt}
\def\small{\@setsize\small{13.6pt}\xipt\@xipt
\abovedisplayskip 13pt plus3pt minus3pt%
\belowdisplayskip \abovedisplayskip
\abovedisplayshortskip  \z@ plus3pt%
\belowdisplayshortskip  7pt plus3.5pt minus0pt
\def\@listi{\parsep 4.5pt plus 2pt minus 1pt
            \itemsep \parsep
            \topsep 9pt plus 3pt minus 3pt}}

\def\underline#1{\relax\ifmmode\@@underline#1\else
        $\@@underline{\hbox{#1}}$\relax\fi}
\@twosidetrue
\relax

\catcode`@=12

\catcode`\@=11

\def\section{\@startsection{section}{1}{\z@}{3.5ex plus 1ex minus
   .2ex}{2.3ex plus .2ex}{\large\bf}}

\def\ps@headings{\def\@oddfoot{}\def\@evenfoot{}
\def\@oddhead{\hbox{}\hfill
        \makebox[.5\textwidth]{\raggedright\ignorespaces --\thepage{}--
        \hfill }}
\def\@evenhead{\@oddhead}
\def\subsectionmark##1{\markboth{##1}{}}
}

\ps@headings

\catcode`\@=12

\relax

\def\figcap{\section*{Figure Captions\markboth
        {FIGURECAPTIONS}{FIGURECAPTIONS}}\list
        {Fig. \arabic{enumi}:\hfill}{\settowidth\labelwidth{Fig. 999:}
        \leftmargin\labelwidth
        \advance\leftmargin\labelsep\usecounter{enumi}}}
 \relax
\def\tablecap{\section*{Table Captions\markboth
        {TABLECAPTIONS}{TABLECAPTIONS}}\list
        {Table \arabic{enumi}:\hfill}{\settowidth\labelwidth{Table 999:}
        \leftmargin\labelwidth
        \advance\leftmargin\labelsep\usecounter{enumi}}}
 \relax
\def\reflist{\section*{References\markboth
        {REFLIST}{REFLIST}}\list
        {[\arabic{enumi}]\hfill}{\settowidth\labelwidth{[999]}
        \leftmargin\labelwidth
        \advance\leftmargin\labelsep\usecounter{enumi}}}
 \relax

\catcode`\@=11

\def\marginnote#1{}
\newcount\hour
\newcount\minute
\newtoks\amorpm
\hour=\time\divide\hour by60
\minute=\time{\multiply\hour by60 \global\advance\minute by-
\hour}
\edef\standardtime{{\ifnum\hour<12 \global\amorpm={am}%
    \else\global\amorpm={pm}\advance\hour by-12 \fi
    \ifnum\hour=0 \hour=12 \fi
    \number\hour:\ifnum\minute<100\fi\number\minute\the\amorpm}}
\edef\militarytime{\number\hour:\ifnum\minute<100\fi\number\minute}
\def\draftlabel#1{{\@bsphack\if@filesw {\let\thepage\relax
  \xdef\@gtempa{\write\@auxout{\string
    \newlabel{#1}{{\@currentlabel}{\thepage}}}}}\@gtempa
    \if@nobreak \ifvmode\nobreak\fi\fi\fi\@esphack}
     \gdef\@eqnlabel{#1}}
\def\@eqnlabel{}
\def\@vacuum{}
\def\draftmarginnote#1{\marginpar{\raggedright\scriptsize\tt#1}}
\def\draft{\oddsidemargin -.5truein
        \def\@oddfoot{\sl preliminary draft \hfil
        \rm\thepage\hfil\sl\today\quad\militarytime}
        \let\@evenfoot\@oddfoot \overfullrule 3pt
        \let\label=\draftlabel
        \let\marginnote=\draftmarginnote
   
\def\@eqnnum{(\theequation)\rlap{\kern\marginparsep\tt\@eqnlabel}%
\global\let\@eqnlabel\@vacuum}  }
\def\preprint{\twocolumn\sloppy\flushbottom\parindent 1em
        \leftmargini 2em\leftmarginv .5em\leftmarginvi .5em
        \oddsidemargin -.5in    \evensidemargin -.5in
        \columnsep 15mm \footheight 0pt
        \textwidth 250mmin      \topmargin  -.4in
        \headheight 12pt \topskip .4in
        \textheight 175mm
        \footskip 0pt
        
\def\@oddhead{\thepage\hfil\addtocounter{page}{1}\thepage}
        \let\@evenhead\@oddhead \def\@oddfoot{} \def\@evenfoot{} 
}
\def\titlepage{\@restonecolfalse\if@twocolumn\@restonecoltrue\onecolumn
     \else \newpage \fi \thispagestyle{empty}\c@page\z@
        \def\thefootnote{\fnsymbol{footnote}} }
\def\endtitlepage{\if@restonecol\twocolumn \else  \fi
        \def\thefootnote{\arabic{footnote}}
        \setcounter{footnote}{0}}  
\catcode`@=12
\relax

\def\ps@headings{\def\@oddfoot{}\def\@evenfoot{}
\def\@oddhead{\hbox{}\hfill
        \makebox[.5\textwidth]{\raggedright\ignorespaces --\thepage{}--
        \hfill }}
\def\@evenhead{\@oddhead}
\def\subsectionmark##1{\markboth{##1}{}}
}

\ps@headings

\relax

\def\firstpage#1#2#3#4#5#6{
\begin{document}
\begin{titlepage}
\nopagebreak
\title{\begin{flushright}
        \vspace*{-1.8in}
        {\normalsize March 1999}\\[-9mm]
        {\normalsize hep-th/9903256}\\[4mm]
\end{flushright}
\vspace{2cm}
{#3}}
\author{\large #4 \\[0.0cm] #5}
\maketitle
\vskip 5mm
\nopagebreak 
\begin{abstract}
{\noindent #6}
\end{abstract}
\vfill
\begin{flushleft}
\rule{16.1cm}{0.2mm}\\
$^{\star}$ {\small Based on talks delivered in September 1998 
at the {\em 32nd 
International Symposium Ahrenshoop on the Theory of Elementary 
Particles}, Buckow;
the {\em 21st Triangular Meeting on Quantum Field Theory}, Crete  
and the TMR meeting on 
{\em Quantum Aspects of Gauge Theories, Supersymmetry and 
Unification}, Corfu; 
to be published 
in the proceedings in Fortschritte der Physik.}   
\\
$^\ast$ {\small e-mail addresses: bakas@nxth04.cern.ch, 
bakas@ajax.physics.upatras.gr} 
\end{flushleft}
\thispagestyle{empty}
\end{titlepage}}

\def\simlt{\stackrel{<}{{}_\sim}}
\def\simgt{\stackrel{>}{{}_\sim}}
\newcommand{\dal}{\raisebox{0.085cm}
{\fbox{\rule{0cm}{0.07cm}\,}}}

\newcommand{\be}{\begin{eqnarray}}
\newcommand{\ee}{\end{eqnarray}}
\newcommand{\btau}{\bar{\tau}}
\newcommand{\p}{\partial}
\newcommand{\bp}{\bar{\partial}}
\renewcommand{\a}{\alpha}
\renewcommand{\b}{\beta}
\newcommand{\g}{\gamma}
\renewcommand{\d}{\delta}
\newcommand{\gsi}{\,\raisebox{-0.13cm}{$\stackrel{\textstyle
>}{\textstyle\sim}$}\,}
\newcommand{\lsi}{\,\raisebox{-0.13cm}{$\stackrel{\textstyle
<}{\textstyle\sim}$}\,}
\date{}
\firstpage{3118}{IC/95/34}
{\large 
{\Large R}EMARKS {\Large O}N {\Large T}HE  
{\Large A}TIYAH - {\Large H}ITCHIN 
{\Large M}ETRIC$^\star$\\
\phantom{X}}
{Ioannis Bakas$^\ast$ 
\phantom{X} }
{\vspace{-.5cm}
\normalsize\sl 
Department of Physics, University of Patras, \\  
\normalsize\sl
GR-26500 Patras, Greece\\} 
{We outline the construction of the Atiyah-Hitchin
metric on the moduli space of $SU(2)$ BPS monopoles with charge 2, 
first as an algebraic curve in $C^3$ following Donaldson and then 
as a solution of the Toda field equations in the continual large $N$
limit. We adopt twistor methods to solve the underlying uniformization  
problem, which by the generalized Legendre transformation yield the
Kahler coordinates and the Kahler potential of the metric. We also comment
on the connection between twistors and the Seiberg-Witten construction
of quantum moduli spaces, as they arise in three dimensional supersymmetric
gauge theories, and briefly address the uniformization of  
algebraic curves in $C^3$ in the context of large $N$ Toda theory.}
\setcounter{section}{0}
\newpage

The Atiyah-Hitchin space is a four-dimensional hyper-Kahler 
manifold with $SO(3)$ isometry that was introduced long time
ago to describe the moduli space of $SU(2)$ BPS monopoles of
magnetic charge 2 \cite{ahgm, ah}.
In more recent
years, this space and various generalizations thereof were
identified with the full quantum moduli space of $N=4$ 
supersymmetric gauge theories in three dimensions \cite{sw}. 
The purpose of this contribution is to summarize briefly various
results scattered in the literature on the algebro-geometric
properties of the Atiyah-Hitchin space, with emphasis on the 
explicit description of its Kahler structure, and make connections
between them by interpreting the classical twistor construction
of hyper-Kahler manifolds (and the associated spectral curve)
essentially as Seiberg-Witten construction of the geometry of 
non-perturbative quantum moduli spaces. The technical ingredient
is the non-triholomorphic nature of the Killing vector fields that
generate the full isometry group of the Atiyah-Hitchin space. 
In the two approaches that will be outlined in the sequel one
chooses an abelian subgroup $SO(2) \subset SO(3)$ to construct 
the Kahler coordinates of the metric and derive all the relevant
information about it in terms of elliptic integrals. One approach
is Donaldson's description of the moduli space as an algebraic
curve in $C^3$, the other is based on the continual Toda field equation 
that arises in the large $N$ limit of 2-dim conformal field theories.
We find as byproduct that $SU(\infty)$ Toda theory 
arises in the uniformization of algebraic curves in $C^3$, whereas
$SU(2)$ Toda theory (ie Liouville's equation) arises in the uniformization
of algebraic curves in $C^2$ (ie Riemann surfaces), as it is well 
known for more than a century. This aspect certainly
deserves further study and we hope to return to it elsewhere together
with the possible use of twistor techniques in future developments
of non-perurbative strings via M-theory \cite{wi}, where the 
Seiberg-Witten curve has a pivotal role, or F-theory \cite{va}.

Recall Donaldson's description of the moduli space of monopoles
as the space of rational functions of a complex variable, say $W$, 
of degree 2 given in normal form as
\be
F(W) = {a_1 W + a_0 \over W^2 - b_0} ; ~~~~~~ 
{a_0}^2 - b_0 {a_1}^2 \neq 0 , 
\ee
modulo the equivalence $F(W) \sim \lambda F(W)$ for $\lambda \in C^{\star}$
\cite{do}.
Here $a_0$, $a_1$, $b_0$ are complex numbers (moduli). 
Setting the resultant of these maps equal to 1, we obtain an algebraic
curve in $C^3$,  
\be
{a_0}^2 - b_0 {a_1}^2 = 1, 
\ee
which describes the (universal) 
double covering of the Atiyah-Hitchin space; 
alternatively, the Atiyah-Hitchin space is the quotient of the
algebraic curve (2) by the involution $Z_2 : (a_0 , a_1 , b_0) 
\rightarrow (-a_0 , -a_1 , b_0)$. 
Donaldson's parametrization of the
moduli space involves choosing a preferred direction in $R^3$ 
(the space where BPS monopoles actually live) 
and the variable $W$ in $F(W)$ above parametrizes
the 2-plane orthogonal to that direction. This description does not
respect the full rotational symmetry of $R^3$, $SO(3)$, 
which is the complete isometry group of Atiyah-Hitchin space, 
as it amounts to
choosing an abelian subgroup $SO(2) \subset SO(3)$; it selects a
preferred complex structure out of the three available in 
hyper-Kahler geometry, and as such it is more appropriate for 
comparison with the Toda field theory description of the metric, 
where, as we will see later, one 
chooses adapted coordinates to a non-triholomorphic $SO(2)$ 
isometry. 

Next, we consider the uniformization of the curve
${a_0}^2 - b_0 {a_1}^2 = 1$, which is only briefly discussed in \cite{ah}.
For this we start first with the elliptic curve 
\be
{\eta}^2 (\zeta) = K^2 (k) \zeta \left( k k^{\prime} {\zeta}^2 
- 3 ({k^{\prime}}^2 - k^2) \zeta - k k^{\prime} \right) , 
\ee
which can be brought into Weierstrass form
$y^2 = 4x^3 - g_2 x - g_3$    
using the change of variables $\eta = k_1 y$, $\zeta = x + k_2$, where
$4{k_1}^2 = k k^{\prime} K^2 (k)$,  
$k_2 = ({k^{\prime}}^2 - k^2)/3k k^{\prime}$.
The quantity ${\omega}_1 = 4 k_1$ is the real period of the curve,
whereas its period matrix ${\omega}_1 / {\omega}_2$ turns out to  
be $\tau = i K(k^{\prime}) / K(k)$. 
For generic values of $k$ the roots in Weierstrass form are all
distinct and they are ordered increasingly on the real axis as 
$e_3 < e_2 < e_1$ 
with branch cuts running from $e_3$ to $e_2$ and $e_1$ to $\infty$. 
As $k \rightarrow 1$ we have $e_2 \rightarrow e_1$ in which 
case the $b$-cycle of the underlying Riemann surface degenerates and
this corresponds to the Taub-NUT limit of the Atiyah-Hitchin space. 
On the other hand, as $k \rightarrow 0$ we have $e_2 \rightarrow e_3$ 
in which case the $a$-cycle of the Riemann surface degenerates and
this corresponds to a bolt structure in the metric, where the 3-dim   
orbits of the full isometry group $SO(3)$ collapse to a two-sphere 
(more precisely $RP_2 \simeq S^2 / Z_2$, since the first homotopy
group of the moduli space is $Z_2$). 
These two limits are related by the modular transformation
$k \leftrightarrow k^{\prime}$ that exchanges the $a$- and $b$-cycles 
of the Riemann surface.   

Next, it is convenient to ``roll up" the curve (3) by applying an $SU(2)$ 
fractional transformation
\be
\zeta \rightarrow \tilde{\zeta} = 
{\bar{c} \zeta - d \over \bar{d} \zeta + c} , ~~~~~
\eta \rightarrow \tilde{\eta} = {\eta \over (\bar{d} \zeta + c)^2} ; ~~~~~
c \bar{c} + d \bar{d} = 1 , 
\ee
where the elements $c$, $d$ of $SU(2)$ are parametrized using Euler angles
$\theta$, $\phi$ and $\psi$ as follows,
\begin{eqnarray}
c & = & e^{{i \over 2} \phi} \left( \sqrt{{1-k \over 2}} {\rm sin} 
{\theta \over 2} e^{-{i \over 2} \psi} - i \sqrt{{1 + k \over 2}} 
{\rm cos} {\theta \over 2} e^{{i \over 2} \psi} \right) ,\nonumber \\ 
d & = & e^{-{i \over 2} \phi} \left( - \sqrt{{1+k \over 2}} {\rm sin}
{\theta \over 2} e^{{i \over 2} \psi} + i \sqrt{{1-k \over 2}}
{\rm cos} {\theta \over 2} e^{-{i \over 2} \psi} \right) .  
\end{eqnarray}
Then the curve (3) becomes
\be
{\tilde{\eta}}^2 (\tilde{\zeta}) = z + v \tilde{\zeta} + 
w {\tilde{\zeta}}^2 - \bar{v} {\tilde{\zeta}}^3 
+ \bar{z} {\tilde{\zeta}}^4 , 
\ee
where the coefficients of the quartic polynomial
turn out to be
\begin{eqnarray}
z & = & -{1 \over 4} K^2 (k) e^{-2i \phi} {\rm sin}^2 \theta 
\left(1 + {k^{\prime}}^2 {\rm sinh}^2 \nu \right) , ~ 
v = -{i \over 2} K^2 (k) {\rm sin} 2 \theta e^{-i \phi} 
(1 + {k^{\prime}}^2 {\rm cos} \psi {\rm tan} \theta 
{\rm sinh} \nu ) \nonumber \\
w & = & {1 \over 2} K^2 (k) \left(2 - {k^{\prime}}^2 + 
3 {\rm sin}^2 \theta \left({k^{\prime}}^2 {\rm cos}^2 \psi 
-1 \right) \right) ; ~~~~ {\rm where} ~~ 
\nu = {\rm log} \left( {\rm tan} {\theta \over 2} \right) 
+ i \psi .  
\end{eqnarray}
Note that
the $\phi$ dependence of the coefficients in (6) can be removed
by appropriate rotation of 
$\tilde{\zeta}$ and $\tilde{\eta} (\tilde{\zeta})$, which is
in turn consistent with the isometry 
$SO(2) \subset SO(3)$ that was selected having $\partial / \partial \phi$
as Killing vector field in terms of Euler angles. 
 
The uniformization proceeds by choosing a complex variable $u$ so
that 
\be
{\cal P} (u) = {d \over \bar{c}} - k_2 \equiv 
{d \over \bar{c}} + e_2 , 
\ee
where ${\cal P} (u)$ denotes the Weierstrass function of the
cubic curve $y^2 = 4x^3 -g_2 x -g_3$. 
Then, according to Atiyah and Hitchin \cite{ah}, 
use of twistor methods lead to the following expressions 
for $b_0$, $a_0$, $a_1$,  
\begin{eqnarray}
\sqrt{b_0} & = & {1 \over 2} {\bar{c}}^2 {\omega}_1 {\cal P}^{\prime} 
(u) , \nonumber \\
{\rm log} \left( a_{0} + \sqrt{b_0} a_1 \right) & = &   
{\omega}_1 \left( \zeta (u) - {{\eta}_1 \over {\omega}_1} u \right) 
+ {1 \over 2} \bar{c} \bar{d} {\omega}_1 {\cal P}^{\prime} (u) ,  
\end{eqnarray}
which together with the algebraic condition 
${a_0}^2 - b_0 {a_1}^2 = 1$ complete the job. Here, $\zeta (u)$ 
is the quasi-periodic $\zeta$-function in the theory of Riemann
surfaces, ${\zeta}^{\prime} (u) = - {\cal P} (u)$, with real 
quasi-period ${\eta}_1$, $\zeta (u + {\omega}_1) = \zeta (u) 
+ {\eta}_1$ (not to be confused with the twistor curve variables
$\zeta$ and $\eta (\zeta)$). The expression (9) can be written in
more compact form using appropriate contour integrals of the 
rotated spectral curve (6). First, by the defining properties of the  
Weierstrass ${\cal P}$-function and the choice (8) of the uniformizing
parameter $u$, we find
\be
b_0 = 4 {\tilde{\eta}}^2 (\tilde{\zeta} = 0) \equiv 4z .     
\ee
Second, following the changes of variables that transform the curve
(3) from its Weierstrass form into the fully rotated spectral curve (6), 
we find upon integration on an $a$-cycle,  
\be
{\rm log} \left(a_0 + \sqrt{b_0} a_1 \right) = {1 \over 2} 
\tilde{\eta} (\tilde{\zeta} = 0) \oint_{a} {d \tilde{\zeta} \over 
\tilde{\zeta} \tilde{\eta}} . 
\ee

Note at this point 
that there are two discrete transformations whose effect is 
more easily described in terms of the uniformizing parameter $u$,  
\be
I_1 : u \rightarrow u + {1 \over 2} ({\omega}_1 + {\omega}_2) , ~~~~~ 
I_2 : u \rightarrow u + {1 \over 2} {\omega}_2 ,  
\ee
which fold pairwise the four Weierstrass points onto each other.
The latter are defined to be the zeros of ${\cal P}^{\prime} (u)$ 
and they occur at $u = {\omega}_1 / 2$ with ${\cal P}(u) = e_1$, 
$u = ({\omega}_1 + {\omega}_2)/2$ with ${\cal P}(u) = e_2$, 
$u = {\omega}_2/2$ with ${\cal P}(u) = e_3$ and $u=0$ with 
${\cal P}(u) = \infty$. 
Under the transformations (12) the coefficients $z$, $v$ and $w$
of the spectral curve (6) remain invariant; we also find that 
under each one of them $b_0$ remains invariant. On the other hand,
${\rm log}(a_0 + \sqrt{b_0} a_1)$ gets shifted by $\pi i$ by applying
either $I_1$ or $I_2$, 
because ${\omega}_1 {\eta}_2 - {\omega}_2 {\eta}_1 = 2 \pi i$.  
Therefore both $I_1$ and $I_2$ correspond to the involution
$(a_0 , a_1 , b_0) \rightarrow (-a_0 , -a_1 , b_0)$ 
and so the algebro-geometric description of the Atiyah-Hitchin space as
a curve in $C^3$, ${a_0}^2 - b_{0} {a_1}^2 = 1$, modulo the 
involution $(a_0 , a_1 , b_0) \rightarrow (-a_0 , -a_1 , b_0)$,
amounts to factoring out both of them. In the   
language of Euler angles, $I_1$ exchanges the position of the two
monopoles, thus treating them as identical particles, and results 
to a bolt structure $S^2$ as $k \rightarrow 0$ by restricting the
range of the angle $\psi$ from 0 to $2 \pi$ instead of $4 \pi$, whereas 
$I_2$, which is the discrete remnant of an additional continuous 
triholomorphic isometry
$\psi \rightarrow \psi + {\rm const.}$ that only appears 
asymptotically as $k \rightarrow 1$, is responsible for the
bolt structure $RP_2 \simeq S^2 /Z_2$ that arises upon factorization.

The second approach to the problem is based on Toda field theory 
which results by selecting an abelian subgroup $SO(2)$ from the
full isometry group $SO(3)$ of Atiyah-Hitchin space and using
adapted coordinates to this isometry. Then, since all isometries
are non-triholomorphic, the hyper-Kahler condition for the metric
amounts to a non-linear differential equation in three dimensions
for a single function $\Psi$, namely Toda field equation, whose solution
determines all components of
the metric. More explicitly, consider Kahler coordinates $q$ and $p$ 
with respect to a selected complex structure   
and let ${\cal K}(q , \bar{q} , 
p , \bar{p} )$ be the Kahler potential that depends on them and 
their complex conjugates. By definition of the Kahler potential 
we have $ds^2 = 2 {\cal K}_{,Q^{A} \bar{Q}^{B}} dQ^A d{\bar Q}^B$, where 
$Q^A$ are $q$ or $p$ and the hyper-Kahler condition for the metric reads
as ${\cal K}_{,q \bar{q}} {\cal K}_{,p \bar{p}} - 
{\cal K}_{,q \bar{p}} {\cal K}_{,p \bar{q}} = 1$.
For manifolds with a non-triholomorphic isometry the Kahler potential
depends on the combination $p \bar{p}$ and not on $p$ and $\bar{p}$
separately, or to put it differently there is a Killing vector field
$\xi = i(p \partial_{p} - \bar{p} \partial_{\bar{p}})$ with the
property $\xi {\cal K} = 0$. 
Then, introducing the notation $p \bar{p} \equiv r$, 
whereas $p / \bar{p} \equiv {\rm exp} (i \sigma)$ defines the 
corresponding phase variable,  
we cast the hyper-Kahler condition into the simpler form 
\be
(r{\cal K}_{,r})_{,r} {\cal K}_{,q \bar{q}} - 
r {\cal K}_{,rq} {\cal K}_{,r \bar{q}} = 1 . 
\ee
Furthermore, by introducing a variable $\rho$ conjugate to 
${\rm log}r$ with respect to ${\cal K}$, 
$r{\cal K}_{,r} = \rho$, we may use 
$(q, \bar{q} , \rho)$ as a new set of independent variables to
rewrite the metric and the condition (13) for the (hyper)-Kahler 
manifold. We obtain the following result for the Kahler metric
\be
ds^2 = {1 \over 2} \left({r \over r_{,\rho}}\right) 
\left(d \sigma + {i \over r} (r_{,q} dq - 
r_{,\bar{q}}d \bar{q}) \right)^2 
+ {1 \over 2} \left({r_{,\rho} \over r}\right) (d{\rho}^2 + 
4r dq d\bar{q}) , 
\ee
whereas the hyper-Kahler condition (13) becomes \cite{bf} 
\be
\left( {\rm log}r \right)_{, q \bar{q}} + r_{, \rho \rho} = 0 , 
\ee
which is the Toda field equation for $SU(N)$ in the continual 
large $N$ limit \cite{sa, ba1}. 

The local coordinate system (14) is called the Toda frame and 
naturally the function $r = {\rm exp} \Psi$ is called the Toda
potential of the metric. 
The change of variables $(q , \bar{q}, r) \rightarrow 
(q, \bar{q}, \rho)$ can be viewed as a Legendre transform in the
sense that the function
\be
{\cal F}(q, \bar{q}, \rho) \equiv \rho {\rm log}r - 
{\cal K} (q, \bar{q}, r) 
\ee
is the ``Hamiltonian" for the ``Lagrangian" ${\cal K}$ with ``momentum" 
$\rho$ and ``velocity" variable ${\rm log}r$. According to this 
we have the relation
$\partial {\cal K} / \partial {\rm log}r \equiv r 
{\cal K}_{,r} = \rho$,  
as required. Then, ${\cal F}_{,\rho} = {\rm log}r$ and 
$r = {\rm exp}\Psi$ 
is considered as a function
of $q$, $\bar{q}$ and $\rho$.   
This concludes our general description of the Kahler structure for 
4-dim hyper-Kahler manifolds with a selected non-triholomorphic isometry, 
generated by $\partial / \partial \sigma$, using the Toda theory approach.  
 
For the Atiyah-Hitchin metric we consider the non-triholomorphic 
isometry generated by $\partial / \partial \phi$ and then by 
application of twistor methods one shows that an appropriate choice
for ${\cal F}$ is \cite{ir}
\be
{\cal F} = -{1 \over \pi i} \oint_{0} {d \tilde{\zeta} \over 
{\tilde{\zeta}}^3} 
{\tilde{\eta}}^2(\tilde{\zeta}) + \oint_{a} 
{d \tilde{\zeta} \over {\tilde{\zeta}}^2} \tilde{\eta}(\tilde{\zeta}) .  
\ee
Here, the first term is given by a contour integral around the origin
and the second by a contour integral over the $a$-cycle of the 
Riemann surface for the twistor space curve (6), which is the same 
as in Donaldson's description of the moduli space. In this context,
${\cal F}$ is a function of the variables $z$, $v$ 
and $w$ given by eq. (7), 
but it follows that it can be used as a generating function for the
Kahler coordinate $q$.
Namely, we have
\be
q = {\partial {\cal F} \over \partial v} = {1 \over 2} 
\oint_{a} {d \tilde{\zeta} \over \tilde{\zeta} \tilde{\eta}} , 
\ee
whereas ${\cal F}_{,w} = 0$ follows by inspection. 
As for the other Kahler coordinate $p$ we have
\be
p = {\tilde{\eta}}^2 (\tilde{\zeta} = 0) = -{1 \over 4} K^2 (k) 
e^{-2i \phi} {\rm sin}^2 \theta \left( 1 + {k^{\prime}}^2 
{\rm sinh}^2 \nu \right)   
\ee
and so $-2 \phi + {\rm arg}(1 + {k^{\prime}}^2 {\rm sinh}^2 \nu)$ 
provides the (shifted) Killing coordinate $\sigma / 2$.  
 
The construction of Kahler coordinates $q$, $p$ by twistor methods
provide a highly non-trivial transcendental solution of the underlying 
continual Toda field equation. The choice (17) for ${\cal F}$ is such
that it remains inert under  
the $SO(2)$ rotations $U= {\rm exp}(i \phi)$ taking 
$\tilde{\zeta} \rightarrow U \tilde{\zeta}$ and 
$\tilde{\eta} \rightarrow U \tilde{\eta}$. In turn, the variables
$q$ and $p$ given by eqs. (18) and (19) transform as follows,  
$q \rightarrow U^{-1} q$ and $p \rightarrow U^2 p$.  
We note at this point that there is another
pair of Kahler coordinates, since their choice is not unique,  
\be
\tilde{p} = 2 \sqrt{p} \equiv \sqrt{b_0} , ~~~~~ 
\tilde{q} = q \sqrt{p} \equiv {\rm log} 
\left(a_0 + \sqrt{b_0} a_1 \right) ,   
\ee
which is preferable in the sense that the $SO(2)$ rotations leave
$\tilde{q}$ inert, whereas $\tilde{p} \rightarrow U \tilde{p}$, 
and connect directly with the variables of the
curve ${a_0}^2 - b_0 {a_1}^2 = 1$ used in Donaldson's description
(see eqs. (10), (11)). We may now drop the tildes and use 
Donaldson's variables (20) to describe the Kahler coordinates of the
metric (14) and the corresponding Toda field equation (15), viewing 
the pair (18) and (19) only as an intermediate choice for the
Atiyah-Hitchin space. In any case, the standard holomorphic 2-form
on the affine surface in $C^3$,
\be
\Omega = {da_1 \wedge db_0 \over a_0} = 2 dq \wedge dp , 
\ee
is not inert under the $SO(2)$ rotations $U$, since 
$\Omega \rightarrow U \Omega$. As such it coincides with 
$F_1 + i F_2$, where $F_1$ and $F_2$ are the two real Kahler forms
of the space that form a doublet under the selected $SO(2)$ 
isometry; the third one, $F_3$, is a singlet and hence inert 
under $U$. 

We conclude with a number of comments which we plan to address
in detail elsewhere \cite{ba}. 
First, the Atiyah-Hitchin space inherits the full isometry group $SO(3)$
by its very construction, since fractional $SU(2)$ 
transformations (5) with Euler angles $\theta$, $\phi$ and $\psi$ have
been used to roll up the elliptic curve (3) to (6). The four
Weierstrass points of the Riemann surface provide the zeros of the
variable $z$ and it can be verified that they occur at 
${\rm cos}^2 \theta = k^2$ and $\psi = \pi / 2$ or $3 \pi / 2$, 
where (6) reduces back to (3). These points are the {\em free field}
points of the underlying Toda field equation (15) in the sense that
${\rm exp} \Psi$ vanishes there; at these points the Atiyah-Hitchin 
metric develops coordinate singularity in the form (14), since one
has ${\rm det} g = ({\Psi}_{,\rho} {\rm exp} \Psi)^2$ and 
${\Psi}_{,\rho} \neq 0$. Guided by the algebro-geometric construction 
of the full space, starting from the {\em free field} curve (3) and
rolling it up to (6) by fractional $SU(2)$ transformations, 
one hopes to device a free field configuration in the 2-dim 
subspace with coordinates $(q, \bar{q})$ so that the transcendental
solution for the Toda potential of the Atiyah-Hitchin space is
reconstructed out of free fields in a group theoretical fashion, 
as it is always the case with 2-dim Toda field equations (though here 
the structure group is infinite dimensional, $SU(\infty)$). However,
the ordinary free field realization of solutions is nothing else
but a way to sum up the perturbative expansion around the free field
configuration, which is provided by group theory in closed form, thanks 
to the integrability of the underlying non-linear differential equations.  
The correct way to address this question here is to consider solutions of 
the Toda field equations defined on a Riemann surface, as indicated
by the uniformization formulae (9) for the Kahler coordinates, and
not on a 2-dim flat Euclidean space which is only appropriate for
considering the asymptotic form of the metric in the Taub-NUT limit
\cite{ba2}.
Then, the non-perturbative corrections, due to instantons, which turn
the perturbative quantum moduli space from Taub-NUT into Atiyah-Hitchin  
in 3-dim $N=4$ supersymmetric gauge theories \cite{sw}, could be
reinterpreted as toron configurations in the Toda frame. This point
of view is certainly useful for future generalizations to more 
complicated examples of algebraic surfaces in $C^3$ due to 
Dancer \cite{da}. 

Second, the non-perturbative construction of Kahler coordinates (in 
particular $q$) and their generating function ${\cal F}$ by 
twistor methods in hyper-Kahler geometry, where one introduces 
an auxiliary curve $\tilde{\eta} (\tilde{\zeta})$ and then integrates
over non-contractible cycles, is identical in vein with the 
construction of Kahler coordinates in quantum moduli spaces of $N=2$
supersymmetric gauge theories a la Seiberg-Witten. We think that this
analogy should be explored further in view of future applications
of twistors in the non-perturbative formulation of strings via 
M-theory or F-theory, as well as in 10-dim supersymmetric Yang-Mills
theories.

Finally, we should explore further the geometric role of large $N$
Toda theory in the uniformization of algebraic curves in $C^3$
that admit a non-triholomorphic $SO(2)$ action, like the 
Atiyah-Hitchin space and its generalizations thereof. In this way
we hope to gain better understanding of non-perturbative aspects
of string theory, whereas the perturbative formulation of strings,
as we know it in terms of world-sheets, is provided by algebraic
curves in $C^2$ (ie Riemann surfaces) and their uniformization via
Liouville theory. This is also an interesting problem in geometry
in its own right.  
 
\vskip1cm
\noindent
{\bf Acknowledgements}   

\noindent
I thank the organizers for their kind invitation and generous
financial support that made possible my participation to these
stimulating events.

\end{document}